\documentclass[letterpaper,prb,preprint]{revtex4}

\usepackage{amssymb,amsmath,rotate,graphicx}

 \def\ltsima{$\; \buildrel < \over \sim \;$} 
\def\simlt{\lower.5ex\hbox{\ltsima}} 
\def\gtsima{$\; \buildrel > \over \sim \;$} 
\def\simgt{\lower.5ex\hbox{\gtsima}} 

\begin{document} 
\title{Ideal glass transitions in thin films: An energy landscape perspective} 
\author{Thomas M. Truskett$^{*}$  \& Venkat Ganesan}
\affiliation{Department of Chemical Engineering and Institute for
  Theoretical Chemistry, The University of Texas at Austin, Austin, TX 78712.}

\vspace{1cm} 
\begin{abstract}
We introduce a mean-field model for the potential
energy landscape of a thin fluid film confined between parallel 
substrates.  The model predicts how the number of accessible basins on
the energy landscape and, consequently, the film's ideal glass
transition temperature depend on bulk pressure, 
film thickness, and the strength of the fluid-fluid and fluid-substrate 
interactions.  The predictions are in qualitative agreement with 
the experimental trends for 
the kinetic glass transition temperature of thin films, suggesting the
utility of landscape-based approaches for studying the behavior
of confined fluids.
\end{abstract}
\maketitle
Amorphous materials confined to small dimensions play a vital 
role in science and technology.  Examples include biological 
fluids in membranes, oil trapped in porous rocks, lubricants, 
layered composites, and thin resist films used in the fabrication 
of microelectronic devices. Many of these systems exhibit 
large surface-to-volume ratios, and thus their physico-chemical 
properties are influenced by boundary and finite-size effects.$^1$
The thermodynamic manifestations of confinement are diverse, ranging 
from shifts of the bulk phase boundaries to the creation of
new phase transitions.$^{2}$  
Dynamics in thin films can also differ markedly from the corresponding 
bulk materials.  Most notably, the glass transition can shift to either 
higher or lower temperatures upon confinement, depending on the 
nature of the interactions of the fluid and the confining medium.$^{3-7}$

The ability to alter the physical properties of thin films   
by tuning film-substrate interactions represents a tremendous 
opportunity for the design and fabrication of advanced materials.  
However, progress hinges on understanding
processes that occur at microscopic and/or mesoscopic length scales.  
Although theoretical and computational studies continue to
provide fundamental insights, many questions concerning the molecular 
origins of thin-film phenomena remain unresolved.$^{8-13}$  Hence,
the development of a consistent and quantitative framework for
modeling thin films is one of the outstanding challenges in 
the engineering and physical sciences.  

In this Communication, we introduce and probe the utility of 
an ``energy-landscape'' based approach$^{14}$ for describing the properties 
of thin films. Specifically, we develop a simple model for the 
topographical features of the energy
landscape of a film confined between parallel substrates.  
We use this model to explore how the thermodynamic 
ideal glass (IG) transition temperature of the film 
(i.e., the temperature at which its 
configurational entropy vanishes) depends on bulk pressure, film thickness,
and the strength of the fluid-fluid and fluid-substrate interactions.
While the existence of an IG transition that underlies the laboratory 
glass transition of real materials remains an open and debated issue,
it is a concept that has proven to be empirically useful and has
strongly influenced modern thought on the glassy state.$^{13}$ 
We find that the predicted IG transition of our model qualitatively 
reproduces experimental trends for the kinetic glass 
transition of thin films.
  
\section*{FLUID FILM MODEL}
The model that we consider is the soft-sphere/mean-field 
(SSMF) fluid.   Its potential energy exhibits
simple scaling properties, and thus it has served as the focus of
several recent studies 
(albeit, in bulk homogeneous conditions).$^{15,16}$ This fluid 
consists of $N$ spherically-symmetric particles that interact via a 
soft-sphere repulsive pair potential $\epsilon (\sigma_{SS}/r)^{n}$ 
in addition to attractions quantified by a density-dependent 
mean-field form $-a_b \rho$ ($\rho=N/V$ is the number density).
The parameter $n$ determines the steepness of the soft-sphere
repulsion, and it is typically taken to be in the range $8 < n <16$.
In 
the bulk situation, the potential energy per particle 
$\varphi ({\bf s}^N, \rho)$ can be expressed in terms of the scaled 
coordinates of the particles ${\bf s}_i = V^{-1/3}{\bf r}_i  \, (i = 1 \cdots N)$ as
\begin{equation}
\label{vgeq1}
\varphi ({\bf s}^N, \rho) = - a_b \rho + y({\bf{s}}^{N})  \eta_b^{n/3}, 
\end{equation}
where $\eta_b = \pi \sigma_{SS}^3 \rho /6$ is the effective packing
fraction of the molecules, and $y({\bf{s}}^{N})$ in Eq.~(\ref{vgeq1}) has 
the dimensions of energy per particle and is defined as
\begin{equation}
\label{tteq1}
y({\bf{s}}^{N}) \equiv \frac{\epsilon}{N} 
\left(\frac{6}{\pi N} \right)^{n/3} \times \sum_{i=1}^{N-1} \sum_{j=i+1}^N
s_{ij}^{-n}
\end{equation}
with $s_{ij}= |{ \bf s_i}-{ \bf s_j}|$. In the situation where 
two parallel substrates (each with surface area $A$) confine the particle 
centers to a film of volume $V=AL$, both the packing fraction $\eta$ 
and the strength of the fluid-fluid attraction parameter $a$ are
functions of the film thickness $L$, and the above model is modified as 
\begin{equation}
\label{vgeq2}
\varphi ({\bf s}^N, \epsilon_L, \rho) = - a\rho - \Psi + y({\bf{s}}^{N}) 
\eta^{n/3} 
\end{equation}
where $\epsilon_L~=~\sigma_{SS}/L$ is the dimensionless reciprocal film
thickness. Here, $-\Psi$ represents the 
attraction between the fluid particles and the confining substrates, 
incorporated in a mean-field manner. Expressions 
for $a$, $\Psi$, and $\eta$ 
have been derived elsewhere$^{17-19}$ and are given by
\begin{eqnarray}
\label{vgeq3}
a &=& a_b \left[1-\frac{3}{4} \epsilon_L
+ \frac{1}{8} \epsilon_L^3
 \right] \nonumber \\
\Psi &=&  \Psi_0 \left[\epsilon_L 
- \frac{\epsilon_L^3}{(1+\epsilon_L)^2}\right]  \nonumber \\
\eta &=& \frac{\pi \sigma_{SS}^3 \rho}{6} 
\left[1 -\frac{3}{16}\epsilon_L \right]
\end{eqnarray}
The parameter $\Psi_0$ establishes the energy scale for the fluid-substrate 
interactions, and its connection to molecular parameters has been discussed 
previously.$^{17,18}$ 

\section*{POTENTIAL ENERGY LANDSCAPES OF THIN FILMS}
For given values of $\rho$ and $\epsilon_L$, $\varphi$ can be 
represented as a hypersurface 
in a $3N+1$ dimensional space --- the film's {\em potential energy 
landscape}.  Despite the multidimensional nature of a material's energy 
landscape, only a few generic features of its ``rugged'' topography 
have been speculated to influence the thermodynamics and dynamics 
of fluids.$^{14-16}$  Here, we develop a simple strategy to
account for how confinement can impact these features.  Since our primary
focus is understanding the behavior of amorphous films, we restrict 
our attention to particle configurations devoid of 
crystalline order.   

(i) {\em Basin enumeration function} $\sigma$: This function 
quantifies the number of distinguishable minima on 
the landscape ({\em inherent structures}).  Explicitly, if 
the total number of (amorphous) inherent structures on 
the landscape with well depths between 
$\phi$ and $\phi + d \phi$ is denoted $d \Omega$, then 
$\sigma(\phi,\epsilon_L,\rho)$ is given by the relationship 
$d \Omega = C \exp \left[N \sigma 
(\phi,\epsilon_L,\rho)\right] d\phi$, where $C$ is a scale factor 
with dimension reciprocal energy.  One of the most simple and
commonly used approximations for the distribution of inherent 
structure depths in bulk fluids is the Gaussian function.$^{16}$
For the case of the SSMF fluid film, we propose 
the following simple generalization:
\begin{equation}
\frac{\sigma(\phi,\epsilon_L,\rho)}{\sigma_{\infty}} = 1- 
\left[ \frac{\phi-\phi_{\infty}}
{\phi_{\infty}-\phi_{m}} \right]^2,\; \phi_{m}< \phi < \phi_{\infty} 
\label{vgeq4}
\end{equation}
where $\phi_{m}\equiv \phi_m (\epsilon_L, \rho)$ and 
$\phi_{\infty} \equiv \phi_\infty(\epsilon_L, \rho)$ are the 
potential energy at the minimum and maximum values of 
the basin enumeration function, respectively. The above expression 
retains the functional form proposed for bulk fluids, while 
rendering the quantities $\phi_\infty$ and $\phi_{m}$ 
film thickness ($\epsilon_L$) dependent.  Unfortunately, the 
accuracy of this approximation is unclear at present because the landscape
statistics of thin films have not been systematically investigated 
by molecular simulation.  Nonetheless, we take~(\ref{vgeq4}) as
a workable starting point, and we leave exploration of more accurate
approximations for future studies. 

(ii) {\em Mean inherent structure energy} $\phi^{*}$:  At a
given (reciprocal) temperature $\beta=1/k_B T$, the
fluid film will spend an overwhelming majority of its time 
in basins of attraction with inherent structures of energy 
$\phi^*=\phi^{*} (\epsilon_L, \rho, \beta)$, which satisfies$^{15,16}$
\begin{equation}
\label{vgeq4_2}
\left( \frac{\partial \sigma}{\partial \phi} \right)_{\epsilon_{L},\rho}
\left[\phi=\phi^*,\epsilon_L,\rho\right]= \beta. 
\end{equation}
Eq.~(\ref{vgeq4_2}) is exact$^{15,16}$ if the intra-basin vibrational 
contribution to the free energy $f_{vib}$ (see below) is independent of 
basin depth $\phi$. 
For the Gaussian landscape (\ref{vgeq4}), we
have
\begin{equation}
\phi^{*} (\epsilon_L, \rho, \beta) =
\begin{cases}
\phi_{\infty}
-\beta {[\phi_{\infty} - \phi_{m}]^2}/{2 \sigma_{\infty}} & \beta <
\beta^{IG} \\ 
\phi_{m}  & \beta \geq \beta^{IG}
\end{cases}
\label{vgeq6}
\end{equation}
where $\beta^{IG} = \beta^{IG} (\epsilon_L, \rho)$ locates the IG
transition. 

(iii) {\em Configurational entropy $s_C$}: 
This quantity is defined as $s_{C} \equiv k_B 
\sigma (\phi^*,\epsilon_L, \rho)$ and thus is given by
\begin{equation}
\label{vgeq8}
\frac{s_{C} (\epsilon_L, \rho, \beta)}{k_B \sigma_{\infty}}  =
\begin{cases}
1 - [{\beta
  (\phi_{\infty}-\phi_m)}/{2\sigma_{\infty}}]^{2} &   \beta <
  \beta^{IG} \\
0 &   \beta \geq \beta^{IG}
\end{cases} 
\end{equation}

(iv) {\em Ideal glass (IG) transition locus} 
$\beta^{IG}$:  At temperatures below the IG transition, 
the configurational entropy of the film vanishes $s_C=0$,
and the system is trapped in the amorphous basin with 
the lowest energy $\phi^*=\phi_m$:
\begin{equation}
\beta^{IG} (\epsilon_L, \rho) = \frac{2
  \sigma_{\infty}}{\phi_{\infty}-\phi_{m}} 
\label{vgeq7}
\end{equation}

(v) {\em Helmholtz free energy} $f$: In terms of
the above quantities, the film's Helmholtz free energy possesses
a simple form:$^{14-16}$
\begin{equation}
\label{vgeq8_2}
f (\epsilon_L, \rho, \beta) = \phi^* - s_C/\beta k_B + f_{vib} 
\end{equation}
\section*{IDEAL GLASS TRANSITION OF THE SSMF FLUID FILM}
Our development in the previous section briefly generalized the
energy landscape formalism for bulk fluids$^{14-16}$ to describe 
fluid films.  Now we quote the explicit
form of the above functions for the SSMF fluid film.  

Using Eq.~(\ref{vgeq2}), (\ref{vgeq3}), and
(\ref{vgeq6})-(\ref{vgeq8_2}), 
we obtain 
\begin{equation}
\label{vgeq10}
\beta^{IG} (\epsilon_L, \rho) = \frac{2\sigma_{\infty}}{\left[y_{\infty} -y_{m}\right]\eta^{n/3}}
\end{equation}
and 
\begin{eqnarray}
\label{vgeq11}
f (\epsilon_L, \rho, \beta) &=& K(\beta) + y_{\infty} \eta^{n/3}  
-
\frac{[y_{\infty}-y_{m}]^2}{4\sigma_{\infty}} \beta
\eta^{2n/3} \nonumber \\ 
&+& \frac{n+2}{2\beta} \ln \eta - a \rho - \Psi
\end{eqnarray}
where $y_m$ and $y_{\infty}$ represent $y({\bf{s}}^{N})$ of 
Eq.~(\ref{tteq1}) evaluated 
at $\phi_m$ and $\phi_{\infty}$, respectively.  The term $K(\beta)$
depends only on temperature and does not enter into our present analysis.  
Following Shell {\em et al.},$^{16}$ we have modeled the vibrational 
contribution to the free energy of Eq.~(\ref{vgeq11}) in the classical 
harmonic approximation.$^{20}$ Other
thermodynamic quantities
of interest, like the transverse component of the pressure tensor 
$P_{\|} (\epsilon_L, \rho, \beta)= \rho(\partial f/ \partial \ln \rho)_{\epsilon_L,\beta}$
and the chemical potential 
$\mu(\epsilon_L, \rho, \beta) = f + P_{\|}\rho^{-1}$
of the film, follow from Eq.~(\ref{vgeq11}).

To examine the confinement-induced shift of the IG transition
for the SSMF fluid film, we consider the situation where the film is
in equilibrium with a bulk fluid at pressure $P_b$; hence the ``shift'' 
we refer to is measured relative to the bulk IG transition at 
that pressure. We have determined the IG transition of the film 
numerically as well as by an approximate analytical theory. In 
both approaches, the IG transition of the bulk fluid is first determined 
by the condition $\beta_b^{IG} (P_b) = \beta^{IG} [\epsilon_L=0, 
\rho_b^{IG} (P_b)]$ (see Eq.~(\ref{vgeq10})), 
where the bulk density $\rho_b^{IG} (P_b)=\rho_b (P_b, \beta_b^{IG})$
follows from an inversion of the equation of state 
$P_{b} = P_{\|} (\epsilon_L=0, \rho_b, 
\beta_b^{IG})$. Similarly, the IG transition of the thin film $\beta^{IG}
(\epsilon_L, P_b)= \beta^{IG} [\epsilon_L,\rho^{IG}(\epsilon_L,P_b)]$ 
is given by Eq.~(\ref{vgeq10}), where the density of the film at its IG 
transition is denoted $\rho^{IG}(\epsilon_L,P_b) \equiv 
\rho(\epsilon_L,P_b,\beta^{IG})$ and is determined by the 
condition that the film and the bulk fluid have equal chemical potentials; 
i.e. $\mu[\epsilon_L=0, \rho_b (P_b,\beta^{IG}), \beta^{IG}] = 
\mu [\epsilon_L,\rho,\beta^{IG}]$.  
Numerical calculations
require values for the parameters $n$, $\sigma_{\infty}$, $y_{\infty}$, 
$y_{m}$, $a_b$, and $\Psi_0$. Here, we use $n=12$, 
$\sigma_{\infty}=0.5368$, $y_{\infty}=61.73 \epsilon$, $y_{m}=53.22 
\epsilon$, and $a_b =16.5 \epsilon \sigma^3$ (we examine several values of the 
fluid-substrate attraction parameter $\Psi_0$).  This set of
parameters was chosen because it provides good qualitative agreement
with both the liquid-state thermodynamics$^{16}$ and the predicted IG 
transition locus$^{21}$ of the bulk Lennard-Jones system. 

A perturbation approach allows us to derive the shifts in the IG 
transition temperature $\Theta \equiv [\beta_b^{IG}(P_b) -
\beta^{IG}(\epsilon_L, P_b)]/\beta^{IG}(\epsilon_L,P_b)$ 
to the first order in $\epsilon_L$. To maintain brevity, we avoid 
elaborating the algebraic details that accompany these calculations, 
and instead quote the following three equivalent results:
\begin{eqnarray}
\Theta & \approx&  \frac{1}{ \Delta c_{P,b}^{IG}}\left[-\left(\frac{\partial s_C} 
{\partial \epsilon_L} \right)_{P_{b}}\right]_{\epsilon_L=0} \epsilon_L
  \nonumber \\
\Theta &\approx&\frac{ 2 \Gamma_b^{IG}/ \{ \sigma_{SS} \rho_b^{IG} \} -3/16} 
{\kappa_{T,b}^{IG}} \left(-\frac{d \ln \beta_b^{IG}}{d P_b}
\right) \epsilon_L \label{vgeq12}\\
\Theta&\approx& \left(\Psi_0 - \frac{9 a_b \rho_b^{IG}}{8} \right)\rho_b^{IG} \left(-\frac{d \ln \beta_b^{IG}}{d P_b}
\right)\epsilon_L
  \nonumber
\end{eqnarray}
Here, $\Delta c_{P,b}^{IG} (P_b)= -(\partial s_C/\partial \ln
\beta)_{P_b}$ is the configurational heat capacity,
$\kappa_{T,b}^{IG}(P_b)$ is the isothermal compressibility,
$2\Gamma_b^{IG} (P_b)/\sigma_{SS}= [\rho^{IG}(\epsilon_L,P_b) - \rho_b
(P_b, \beta^{IG})]/\epsilon_L$ is the surface excess density, and $\eta_b^{IG}
(P_b)$ is the packing fraction, each evaluated at the IG transition of the
bulk fluid, i.e. $[\epsilon_L=0,P_b,\beta_b^{IG} (P_b)]$.  The
thermodynamic quantities in (13) can be expressed$^{22}$ as 
analytical functions of the molecular parameters of the model.
Figure 1 
displays the confinement-induced shifts in the IG transition 
temperature (whose physical implications are discussed below), 
demonstrating the excellent accuracy of the perturbation approach in 
capturing the full numerical results.  

The above expressions (\ref{vgeq12}) shed light on the physics of 
confinement-induced changes in the IG transition temperature. 
Consider the first equality of Eq.~(\ref{vgeq12}). Since 
$\Delta c_{P,b}^{IG} (P_b)>0$, this relation reveals
the following simple rule for the (small $\epsilon_L$) shift of the IG
transition.  If confinement increases the number of basins 
on the landscape that the fluid can sample 
(hence increasing $s_C)$, then the IG transition temperature is
depressed.  Conversely, if confinement decreases the number of 
accessible basins, then the IG transition temperature is elevated.
This result does not depend upon a specific model for the film's 
energy landscape.

The second equality of Eq.~(\ref{vgeq12}) uses our model to connect the 
landscape-based perspective to physical quantities.  Since 
$d \ln \beta^{IG}/d P_b < 0$ for this model, it predicts
that the direction of the shift of the IG transition 
is determined by the sign and magnitude of the film's surface excess 
density $2\Gamma_b^{IG} (P_b)$.  Large positive values of the surface excess 
density reduce the configurational entropy of the film
and increase the IG transition temperature, while small or negative
values of the surface excess have the opposite effect.  While this 
result depends on our model for the energy landscape, we note that it 
is consistent with the recent theoretical predictions 
of McCoy and Curro.$^{12}$  The reason that these two different 
approaches arrive at similar conclusions is easy to understand.  McCoy and 
Curro's model begins with the hypothesis that the confinement-induced
shift in the {\em kinetic} glass transition is 
determined by how confinement affects the density of the fluid film.  
Similarly, as can be seen by Eq.~(\ref{vgeq10}), the shift in the IG transition of our 
mean-field model is determined by how confinement affects the packing 
fraction of the molecules in the film.     

The third equality of Eq.~(\ref{vgeq12}) establishes a molecular connection:  strongly
attracting walls $\Psi_0 \sigma_{SS}^3/a_b >> 1$ elevate the IG
transition while neutral or repulsive walls 
$\Psi_0 \sigma_{SS}^3/a_b \approx 0$ depress the IG transition.  
Fig.~1 shows the confinement-induced shifts in the IG 
transition temperature as predicted by the linear expression of
Eq.~(13) and the full non-linear model of Eq.~(11) and (12).  
As can be seen, the shift
of the IG transition is approximately inversely proportional to 
film thickness down to molecular length scales.         
As has been discussed extensively elsewhere, these trends are in
good qualitative agreement with the experimental shifts in the 
glass transition temperature of confined fluids.$^{3,4,6}$

Finally, we note that the glass transition shifts shown in
Fig.~1 correspond to a fixed value of the bulk pressure $P_b=0$.  
It is straightforward to employ the model and the framework outlined 
above to analyze the effects of varying the bulk pressure. 
Preliminary calculations indicate that while qualitatively similar results 
are seen at higher pressures, reducing the pressure (placing the fluid 
under tension) can change both the sign and the magnitude of 
shifts.$^{22}$ A comprehensive investigation of pressure
effects on the glass transition is beyond the scope of this 
Communication and is deferred to a future publication.  

\section*{CONCLUSIONS}
We have examined the confinement-induced changes 
in the energy landscape of a simple statistical mechanical model 
of a fluid film. Our analysis here focuses on elucidating how
the film's IG transition depends on various molecular and 
macroscopic parameters. The model predictions are
qualitatively consistent with theoretical and experimental studies, 
thereby suggesting that landscape based approaches
may serve to provide a framework for predicting the thermodynamic
and dynamic properties of confined fluids. It is appropriate to point 
out that the SSMF film model, although conceptually
useful, should only be viewed as a starting point that suggests 
the viability of the inherent structure 
formalism$^{14}$ for understanding the behavior of confined
fluids and thin films. 
In fact, many films of scientific interest exhibit either narrow confining 
geometries and/or strong, directional fluid-substrate attractions 
that are not amenable to a mean-field treatment or the Gaussian
landscape approximation.  
In such cases, molecular simulations can be employed to
extract information about how confinement alters their energy
landscapes, and hence their physico-chemical properties.  
We are currently pursuing research along these 
lines.   

We thank Scott Shell, Pablo Debenedetti, Emilia La Nave and 
Francesco Sciortino for sending us a preprint of their
recent manuscript$^{16}$, which played an important role in motivating the
present work. VG gratefully acknowledges support from 
National Science Foundation under Award Number DMR-02-04199, and the 
Petroleum Research Fund, administered by the ACS. 

\vspace{0.2in}

\centerline{{\bf{REFERENCES}}} 

\parindent=0truecm \hangindent=0.6truecm \hangafter=1
$^*$Author to whom correspondence should be addressed.

\parindent=0truecm \hangindent=0.6truecm \hangafter=1
$^1$J. M. Drake and J. Klafter, Phys. Today {\bf 43}, 46-55 (1990).

\parindent=0truecm \hangindent=0.6truecm \hangafter=1
$^2$L.~D. Gelb, K.~E. Gubbins, R. Radhakrishnan, and M. 
Sliwinska-Bartkowiak, Rep. Prog. Phys. {\bf 62}, 1573 (1999);

\parindent=0truecm \hangindent=0.6truecm \hangafter=1
$^3$J. A. Forrest and K. Dalnoki-Veress, Adv. Colloid Interface
Sci. {\bf 94}, 167 (2001).

\parindent=0truecm \hangindent=0.6truecm \hangafter=1
$^4$G.B. McKenna, Journal de Physique IV (France) {\bf 10}, Pr7-53
(2000).

\parindent=0truecm \hangindent=0.6truecm \hangafter=1
$^5$J. Sch\"{u}ller, R. Richert, and E.~W. Fischer, Phys. Rev. B
{\bf 52}, 15232 (1995). 

\parindent=0truecm \hangindent=0.6truecm \hangafter=1
$^6$D.~S. Fryer, P.~F. Nealey, J.~J. de Pablo, Macromolecules 
{\bf 33}, 3376 (2000); D.~S. Fryer {\em et al.}, Macromolecules
{\bf 34}, 5627 (2001); R.~S. Tate {\em et. al.}, J. Chem. Phys.
{\bf 115}, 9982 (2001).

\parindent=0truecm \hangindent=0.6truecm \hangafter=1
$^7$R.~A.~L. Jones, Curr. Opin. Colloid Interface Sci. {\bf 4}, 153
(1999).

\parindent=0truecm \hangindent=0.6truecm \hangafter=1
$^8$C.~Mischler, J. Baschnagel, and K. Binder, Adv. Colloid Interface
Sci. {\bf 94}, 197 (2001).

\parindent=0truecm \hangindent=0.6truecm \hangafter=1
$^9$ J.~A. Torres, P.~F. Nealey, and J.~J. de Pablo, Phys. Rev. Lett.
{\bf 85}, 3221 (2000).

\parindent=0truecm \hangindent=0.6truecm \hangafter=1
$^{10}$ T~R. B\"{o}hme and J.~J. de Pablo, J. Chem. Phys. {\bf 116}, 9939
(2002).

\parindent=0truecm \hangindent=0.6truecm \hangafter=1
$^{11}$ P.~G. de Gennes, Eur. Phys. J. E {\bf 2}, 201 (2000); D.~Long 
and F.~Lequeux, Eur. Phys. J. E {\bf 4}, 371 (2001).

\parindent=0truecm \hangindent=0.6truecm \hangafter=1
$^{12}$J.~D. McCoy and J.~G. Curro, J. Chem. Phys. 
{\bf 116}, 9154 (2002).

\parindent=0truecm \hangindent=0.6truecm \hangafter=1
$^{13}$ W. Kauzmann, Chem. Rev. {\bf 43}, 219 (1948); 
J.~H. Gibbs and E.~A. Dimarzio, J. Chem. Phys. {\bf 28}, 373 
(1958);
T.~R. Kirkpatrick and P.~G. Wolynes, Phys. Rev. B {\bf 36}, 8552
(1987); F.~H. Stillinger, J. Chem. Phys. {\bf 88}, 7818 (1988);
M. M\'{e}zard and G. Parisi, Phys. Rev. Lett. {\bf 82}, 747 (1999).

\parindent=0truecm \hangindent=0.6truecm \hangafter=1
$^{14}$F.~H. Stillinger and T.~A. Weber, Phys. Rev. A {\bf 25}, 978 
(1982).

\parindent=0truecm \hangindent=0.6truecm \hangafter=1
$^{15}$P.~G. Debenedetti, F.~H. Stillinger, T.~M. Truskett, and 
C.~J. Roberts, J. Phys. Chem. B. {\bf 103} (1999); P.~G. Debenedetti,
T.~M. Truskett, C.~P. Lewis, and F.~H. Stillinger,
Adv. Chem. Eng. {\bf 28}, 21 (2001); F.~H. Stillinger,
P.~G. Debenedetti, and T.~M. Truskett {\bf 105}, 11809 (2001).

\parindent=0truecm \hangindent=0.6truecm \hangafter=1
$^{16}$M.~S. Shell, P.~G. Debenedetti, E. La Nave, and F. Sciortino,
J. Chem. Phys {\bf 118}, 8821 (2003). 

\parindent=0truecm \hangindent=0.6truecm \hangafter=1
$^{17}$M. Schoen and D.~J. Diestler, J. Chem. Phys. {\bf 109}, 5596 
(1998).

\parindent=0truecm \hangindent=0.6truecm \hangafter=1
$^{18}$T.~M. Truskett, P.~G. Debenedetti, and S. Torquato, 
J. Chem. Phys. {\bf 114} 2401 (2001).

\parindent=0truecm \hangindent=0.6truecm \hangafter=1
$^{19}$ The mean-field expressions given in Eq.~(\ref{vgeq3})
represent the case 
where the effective ``hard-core'' diameters of the fluid-fluid and 
fluid-substrate interactions are equal.

\parindent=0truecm \hangindent=0.6truecm \hangafter=1
$^{20}$A.~Scala, F.~W. Starr, E. La Nave, F. Sciortino, and H.~E.
Stanley, Nature {\bf 406}, 166 (2000); S. Sastry, Nature {\bf 409}, 
164 (2001); I. Saika-Voivod, P.~H. Poole, and F.~Sciortino, 
Nature {\bf 412}, 514 (2001). 

\parindent=0truecm \hangindent=0.6truecm \hangafter=1
$^{21}$R. Di Leonardo, L. Angelani, G. Parisi, and 
G. Ruocco, Phys. Rev. Lett. {\bf 84}, 6054 (2000).

\parindent=0truecm \hangindent=0.6truecm \hangafter=1
$^{22}$T.~M. Truskett and V. Ganesan, unpublished.

\newpage

{\bf FIGURE CAPTIONS} \\

{\noindent {FIG. 1. The confinement induced shift in the IG 
transition $\Theta$ as a function of $\epsilon_L$ for the SSMF fluid 
film.  The solid lines represent the approximation of Eq.~(\ref{vgeq12}) 
and the symbols are the predictions of the full nonlinear model of 
Eq.~(\ref{vgeq10}) and (\ref{vgeq11}).  The relative strength of
 the fluid-substrate to 
fluid-fluid attractions is quantified by 
$\Psi_0 \sigma_{SS}^3/a_b$. The above results correspond to the case wherein the films are in equilibrium with the bulk SSMF fluid at $P_b=0$.

\newpage
\vfill
\begin{figure}[h]
\centering
\rotatebox{-90}{\includegraphics[bb= 113 111 539 573,height=4in]{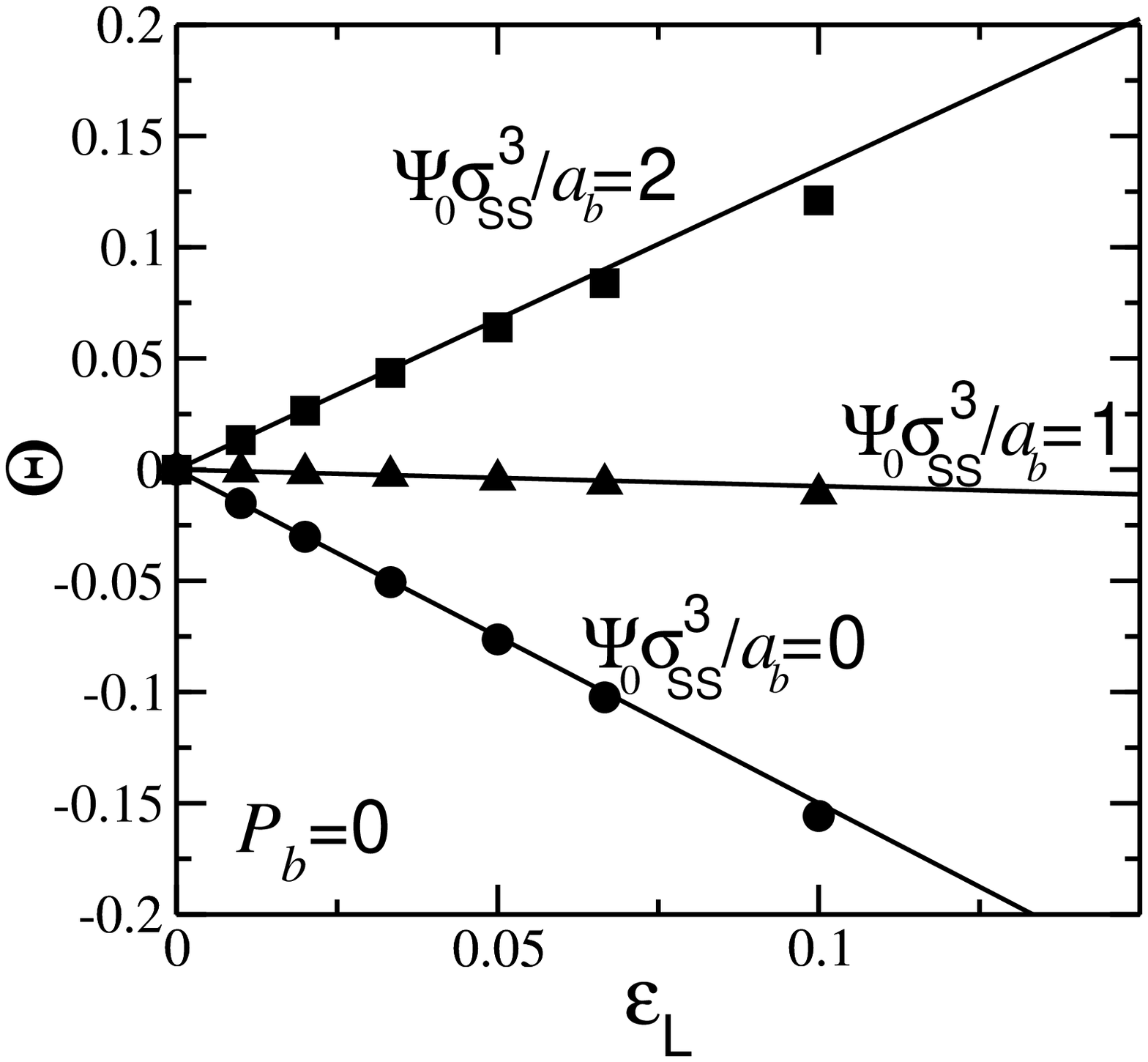}}
\label{fig1}
\end{figure}
\vfill
\centerline{\large{FIG. 1 -- Truskett and Ganesan, 2003}}

\end{document}